# The ASTRI Cherenkov Camera: from the prototype to the industrial version for the Mini-Array

G. Sottile[*], P. Sangiorgi[*], C. Gargano[*], F. Lo Gerfo[*], M. Corpora[*], O. Catalano[*], D. Impiombato[*], D. Mollica[*], M. Capalbi[*], T. Mineo[*], G. Contino[*], B. Biondo[*], F. Russo[*], M. C. Maccarone[*], G. La Rosa[*], S. Giarrusso[*], G. Leto[†], A. Grillo[†], G. Bonanno[†], G. Romeo[†], S. Garozzo[†], D. Marano[†], V. Conforti[‡], F. Gianotti[‡], S. Scuderi[§], G. Pareschi[¶], G. Tosti[∥], A. Abba[**], A. Cusimano[**], F. Caponio[**], C. Tintori[††], M. Lippi[††], F. Vivaldi[††], G. Marchiori[‡‡], M. Spinola[‡‡], A. Colovini[‡‡], F. Perez[x], S. Ahmad[x], J. B. Cizel[x], J. Fluery[x]

*for the ASTRI project.*

[*] - INAF IASF Palermo, Italy
[†] - INAF OACT Catania, Italy
[‡] - INAF OAS Bologna, Italy
[§] - INAF IASF Milano, Italy
[¶] - INAF OA Brera Milano, Italy
[∥] - Università degli Studi di Perugia, Italy
[**] - Nuclear Instruments - Lambrugo (CO), Italy
[††] - CAEN S.p.A. - Viareggio (LU), Italy
[‡‡] - EIE Group s.r.l. - Venezia, Italy
[x] - Weeroc - Villebone sur Yvette, France

## Abstract

The observation of energetic astronomical sources emitting very high-energy gamma-rays in the TeV spectral range (as e.g. supernova remnants or blazars) is mainly based on detecting the Cherenkov light induced by relativistic particles in the showers produced by the photon interaction with the Earth atmosphere. The ASTRI Mini-Array is an INAF-led project aimed observing such celestial objects in the 1 - 100 TeV energy range. It consists of an array of nine innovative imaging atmospheric Cherenkov telescopes that are an evolution of the dual-mirror aplanatic ASTRI-Horn telescope operating at the INAF "M.C. Fracastoro" observing station (Serra La Nave, Mount Etna, Italy). The ASTRI Mini-Array is currently under construction at the Observatorio del Teide (Tenerife, Spain). In this paper, we present the compact (diameter 660mm, height 520mm, weight 73kg) ASTRI-Horn prototype Cherenkov Camera based on a modular multipixel Silicon Photon Multiplier (SiPM) detector, has been acquiring data since 2016 and allowing us to obtain both scientific data and essential lessons. In this contribution, we report the main features of the camera and its evolution toward the new Cherenkov camera, which will be installed on each ASTRI Mini-Array telescope to cover an unprecedented field of view of 10.5°.

## 1. Introduction

ASTRI (*Astrofisica con Specchi a Tecnologia Replicante Italiana*) is a project financed by the Italian Ministry of Education, University and Research (MIUR) and led by the Italian National Institute for Astrophysics (INAF). The objective of the project was to construct an end-to-end prototype for the Small Size Telescopes \cite{pareschi2016astri} of the CTA observatory. The telescope belongs to the Imaging Atmospheric Cherenkov Telescope (IACT) class because its camera detects the Cherenkov light produced by particle showers originating from celestial sources' gamma-ray photons interacting with the Earth's atmosphere. The first result of this project has been the implementation of the ASTRI telescope, operated at the INAF "M.C. Fracastoro" observing station (Serra La Nave at 1725m a.s.l., Mount Etna, Italy, Maccarone et al., 2013). This represents the first dual-mirror Schwarzschild-Couder telescope, whose optical conception was validated in 2014 (Giro et al., 2014). Moreover, ASTRI provided in 2018 the first detection with a dual mirror telescope of a gamma ray source, the CRAB nebula observed at energies > 3 TeV (Lombardi et al., 2020).

The ASTRI telescope, dedicated in 2018 Guido Horn D'Arturo (the Italian-Jewish astronomer who introduced the use of segmented mirrors in astronomy), is based on a primary segmented 4m-diameter mirror, while the secondary one is monolithic with a diameter of 1.8m; both mirrors are aplanatic (Sironi, 2017).

This optical configuration is also, characterized by a curved focal plane (to compensate for astigmatism) and short depth of focus (+/- 1mm), with a small plate scale (37.5 mm/deg). This drove the design of a compact and lightweight camera. The ASTRI-Horn camera (Sottile et al., 2016, Catalano et al., 2013) introduced innovative technology to employ in the field of the IACT. It was one of the first instruments to use the Silicon PhotoMultiplier (SiPM) detectors and to perform signal processing by a combination of pulse shaping and peak detection, controlled by an external signal.

This architecture has advantages compared to the traditional Cherenkov camera working at a very high sampling rate. Although the first tests performed with the ASTRI-Horn camera demonstrated its scientific potential several lessons were learned. The heritage of the camera is now being exploited to develop the first industrial version, which will be realized by CAEN – EIE industrial consortium under INAF's supervision. The plan is to build eleven cameras for ASTRI Mini-Array (Vercellone et al., 2020; Scuderi et al., 2022). The first industrial engineering model will be installed at Teide at the end of April 2023, with the purpose of validating the operation flow of the whole system. After this phase, ten cameras will be built to be mounted at the nine telescopes of the array, while the tenth one will be used as a spare. This paper reports the main sub-systems of the ASTRI-Horn camera, which led to the evolution of the camera used for the ASTRI Mini-Array.

### 2.1 ASTRI camera main sub-systems

The ASTRI-Horn and Mini-Array cameras are composed of the identical subsystems: SiPM, Front End Electronics (FEE), Back End Electronics (BEE), Voltage Distribution Box (VDB), Thermal-Mechanical Assembly and Calibration System.

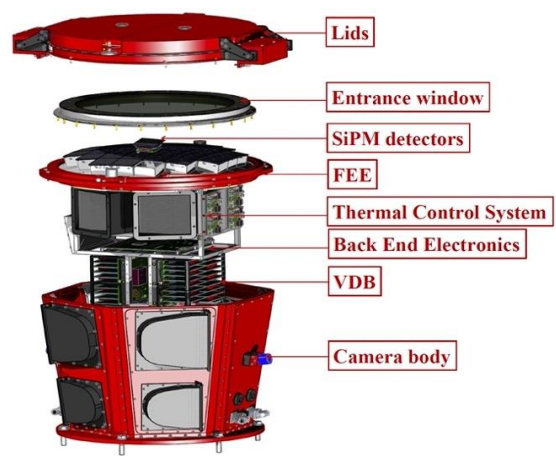

*Fig. 1 Exploded 3D layout of the ASTRI-Horn camera.*

## 2.1 SiPM tile, Front End Electronics and FPGA board

The ASTRI-Horn camera is based on the Hamamatsu Silicon PhotoMultiplier (SiPM) of 7mm x 7mm (S10943), organized in tiles of 8 x 8 pixels. The new camera has the same layout but employs the newer Hamamatsu S14521 photodetectors (Romeo et al., 2018).

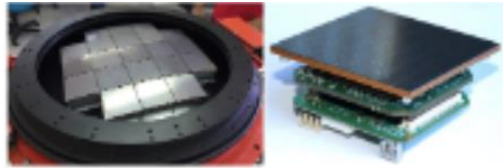

*FIG. 2. ASTRI Mini-Array SiPM tile (56mm x 56mm) - Hamamatsu S14521 75um.*

37 PDMs populate the focal plane, which has a spherical-cap shape with a curvature radius of 1060mm and a diameter of 445mm, to match the focus of the Schwarzschild-Couder configuration of the telescope. Thanks to the 37 PDMs, the camera will comprise a total number of pixels of 2368 with a field of view of 10.5°. In the case of ASTRI-Horn, only 21 PDMs were installed (Fig. 2) due to procurement issues affecting the SiPM detectors. In the case of the ASTRI-Horn camera, the total number of pixels is 1344, with a reduced field of view of 7.8°.

The output signals of the SiPM detectors are processed by two Citiroc ASICs by the Weeroc French company (https//www.weeroc.com, Impiombato et al., 2015) mounted on the ASIC board (Fig. 3). The Citiroc has 32 channels, so two ASICs are required to process the 8 x 8 pixels of the SiPM tile.

Each channel has two processing chains with two possible gain configurations (High Gain - HG and Low Gain - LG), to cover a wide dynamic range between 0 to 2000 Photoelectron Equivalent (PE), per SiPM gain of $1 \times 10^6$. Both chains have a Peak Detector (PD) to measure the amplitude of the signals. The PDs are armed by a Camera Trigger Signal to catch the peaks of the SiPMs signal of the whole focal plane. Then, only the HG and LG peaks are digitally converted by external Analog-to-Digital converters (ADCs).

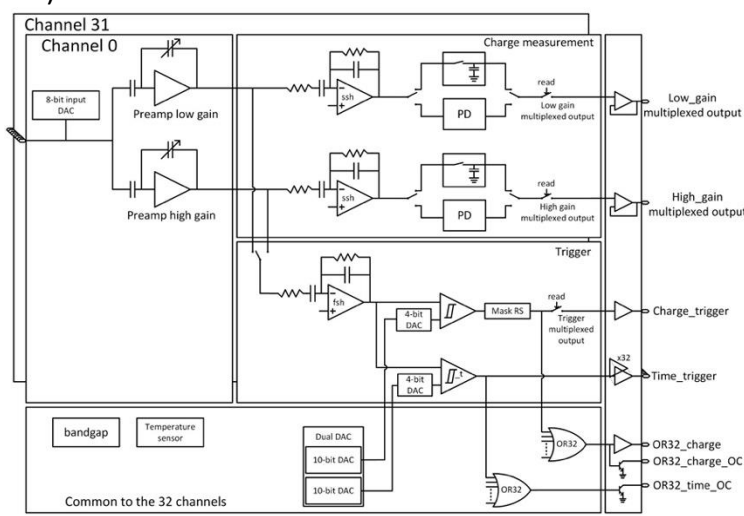

*Fig. 3. Citiroc ASIC block diagram.*

This is the main difference between the ASTRI-Horn camera and the other Cherenkov cameras, which continuously sample the photodetectors signal at a very high rate (Otte et al., 2015; Schoorlemmer et al., 2017; Chernov et al., 2020).

The architecture with PDs produces two main advantages. The first one is the very low amount of data produced per scientific event detected that must be transferred and processed. The second advantage is the low power required by Citiroc to process the signals (300mW/chip). The small amount of heat produced makes the design optimization simpler regarding signal-to-noise ratio because the ASICs can be placed very close to the SiPM, avoiding the risk of warming these temperature-sensitive photodetectors up.

The Citiroc provides a trigger signal through a discriminator per channel with a shared programmable threshold by a 10-bit Digital-to-Analog converter (DAC). To compensate for the non-uniformity of the 32 discriminators due to the typical dispersion of the fabrication technology of the die, the Citiroc ASIC allows us a fine-tuning of the threshold channel-by-channel 4-bit DAC. This feature permits a uniform trigger signal channel-by-channel and makes more effective matching of the two Citiroc ASICs of the same board.

The ASIC board for the ASTRI Mini-Array has been updated in a few details, like the new version of the Citiroc 1A provided in the BGA package and the higher resolution of the ADCs from 12-bit of the old one to 14 bit of the new one.

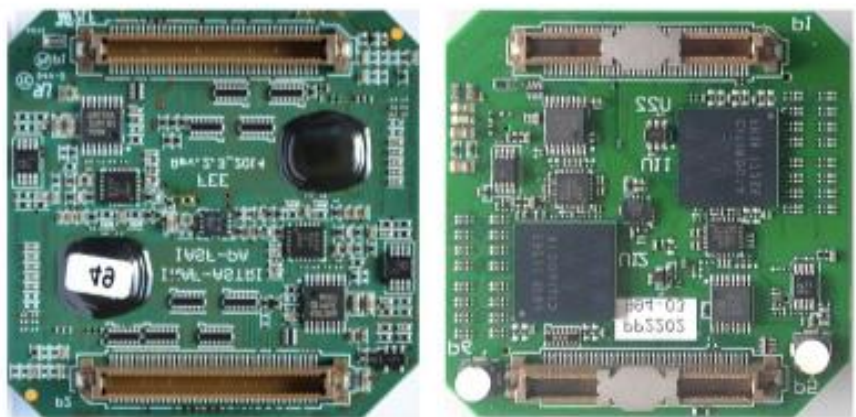

*Fig. 4. Comparison of the ASIC boards (50mm × 50mm); on the left the ASTRI-Horn configuration, on the right the one adopted for the cameras of the ASTRI Mini-Array.*

Based on Xilinx Artix 7 FPGA device, the FPGA board interfaces the ASIC board. The FPGA controls the ASICs configurations and the ADCs to the digital conversion of the Citiroc analogue outputs (high gain and low gain). A custom FPGA firmware was developed to run several algorithms among which the delivering of a topological trigger, activated when the signals of a given number of contiguous pixels are above the 10-bit DAC threshold set at the same time. The topological trigger avoids fake triggers due to stars within the telescope field of view. The trigger signal of the FPGA board, called the PDM trigger, is sent to the FPGA trigger of the BEE, which generates the Camera trigger signal (i.e. the second trigger level) after elaboration. The BEE provides the camera trigger signal to all PDMs of the focal plane. The FPGA manages the camera trigger signal to arm the peak detectors and to capture the amplitude of the event detected. Another essential function running in the FPGA is the elaboration of the signals for the so called "variance" mode (Segreto et al., 2019; Iovenitti et al., 2022). This technique is based on the statistical analysis of the SiPM signal continuously sampled but without giving rise to any trigger. The mean ADC values are constant with time, but the variance value is proportional to the light flux detected by the pixel. The variance technique does not interfere with the Cherenkov event detection because it is performed during the time intervals in which no events are detected. A variance is an essential tool to check the health status of the focal plane, to monitor the alignment of the mirrors and pointing of the telescope (Fig. 5). The FPGA board of the ASTRI Mini-Array differs from the one of the prototype cameras just for a few details, which have increased the reliability. The connectors have been replaced with more robust items that prevent accidental disconnections. Moreover, the power supply has been re-designed with linear regulators to reduce the radiated noise toward the sensitive inputs of the Citiroc ASICs.

### 2.2 Back End Electronics

The BEE board is the computer of the ASTRI camera. The architecture is based on SoC Xilinx Zynq 7000 and Linux OS and other two Xilinx Artix-7 FPGAs. An Artix-7 is dedicated to the camera trigger signal generation. It checks the 37 incoming PDM trigger differential lines to detect a trigger event. Accordingly, the secondary trigger algorithm generates and provides the camera trigger signal to the 37 PDMs. The second Artix-7 FPGA is in charge of receiving the scientific data from the PDMs and to managing the slow control with the PDMs. In the of the ASTRI-Horn prototype camera, the Zynq 7000 FPGA collects the scientific and housekeeping data, creates the packets and sends them to the camera server. The BEE of ASTRI Mini-Array architecture is the improved version of the ASTRI-Horn one. The Zynq 7000 has been replaced with a more powerful Zynq Ultrascale+ installed on Enclustra module. The new BEE, shown in Fig. 4, has been designed so that each FPGA has a dedicated power supply unit. Moreover, now it accommodates more reliable and robust connectors, able to provide shielded communication links.

Beyond the camera trigger signal, elaborated inside the trigger FPGA, the new BEE can generate a faster trigger signal thanks to a wired-or connection. This feature will allow us to set the shortest shaping time of the Citiroc ASICs and, consequently, achieve a better signal-to-noise ratio of the event acquisition. Finally, the ASTRI Mini-Array employs the White Rabbit (WR) technology for sub-nanosecond synchronization among nine telescopes. The new BEE manages the 10MHz and PPS inputs signals of the WR. The 10MHz signal is the fundamental clock from which the PLLs system of the BEE derives the clocks for the synchronism of all sub-systems. The PPS and 10MHz signals allow us to time tag each triggered event.

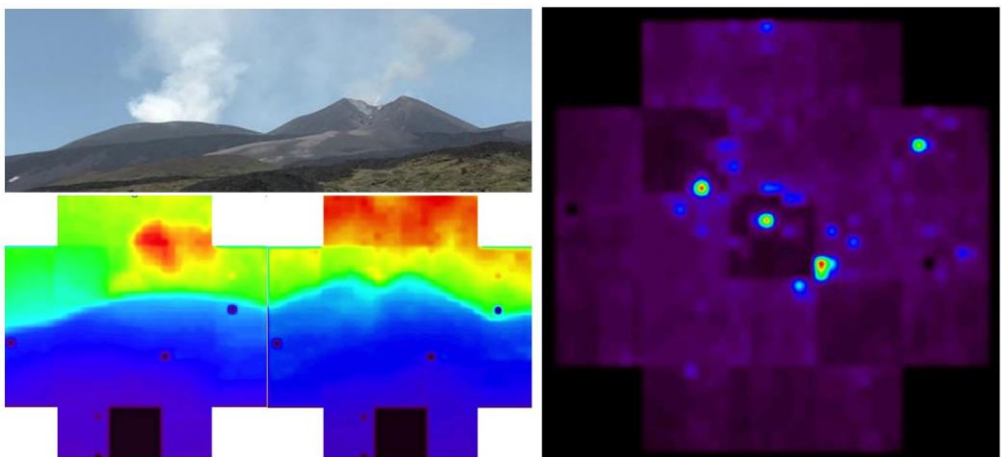

*Fig. 5: The variance method as pointing monitor tool: on the left the Etna volcano profile, on the right the Orion's Belt.*

## 2.3 Voltage Distribution Box

The Voltage Distribution Box is the power supply of the PDMs. It has a modular architecture composed of two identical main boards, each managing 19 daughter boards. One of the main board - identified as master - receives the commands from BEE through a SPI differential bus. It routes the messages to their daughter boards or to the second main board. The main boards communicate with the respective daughterboards through RS485 bus. The VDB of ASTRI Mini-Array has a simpler architecture than the prototype, as shown in Fig. 5, where the main boards are a single point of failure. The master main board failure would cause the loss of the whole focal plane because all the PDMs would be unpowered. The failure of the slave main board would cause the loss of half focal plane. In both cases, the camera wouldn't be able to operate. The main boards of the new VDB are passive just to hold the daughter boards, which are connected to BEE through two I2C buses.

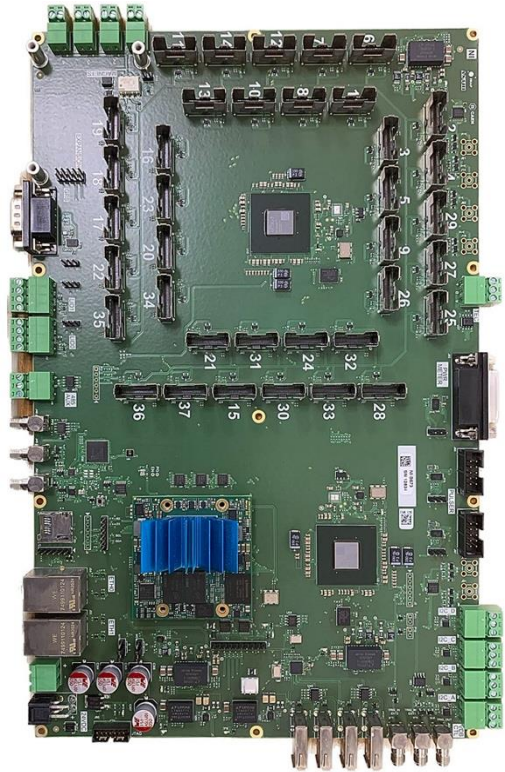

*Fig. 6: ASTRI Mini-Array BEE (200mm × 325mm).*

The new architecture is more reliable and has more straightforward firmware. Another step forward is the high voltage module to supply the SiPMs detectors, capable to provide up to 70mA, (15mA for the ASTRI-Horn). The increased current capability permits sustaining the more considerable gain of the new photodetectors and exposing the focal plane to a higher level of night sky background. This has a positive effect on the duty cycle of the observation; in other words, the telescope will be able to be used for more nights per year. Another essential characteristic is the voltage-temperature compensation of the SiPM bias operated directly by the daughter board rather than being controlled by the BEE.

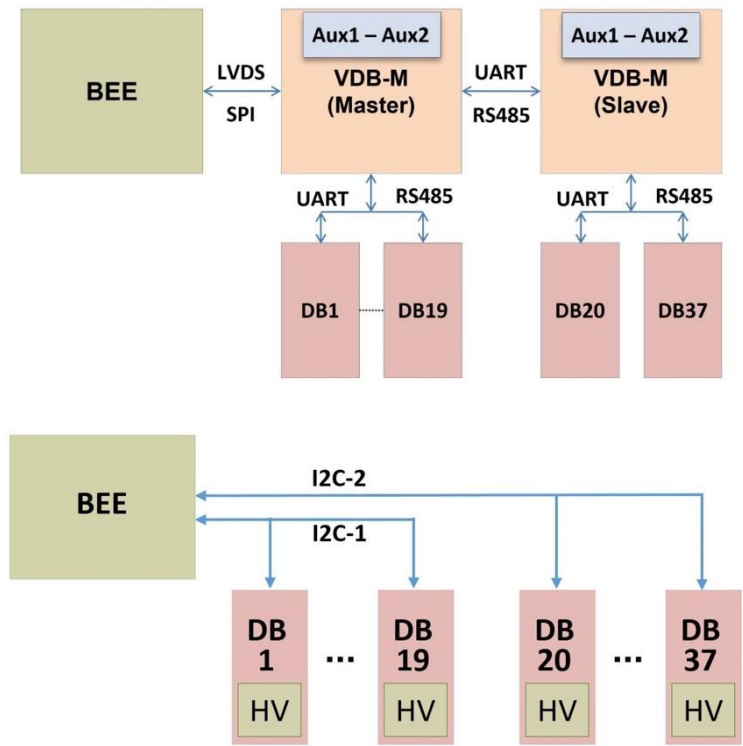

*Fig. 7 VDB architectures for the ASTRI-Horn (above) and Mini-Array (below) cameras.*

### 1. Thermal Control system

The thermal control system is essential in a Cherenkov camera based on SiPMs. The gain and the dark count rate of these photodetectors are very sensitive to temperature. Therefore, it is fundamental to keep equalized and constant temperature over the entire focal plane. The ASTRI camera has a Focal Plane Support Structure (FPSS) made of aluminum with a network of heat pipes embedded in its thickness. Four Thermoelectric Coolers (TEC) can cool down or warm up the FPSS in order to maintain the focal plane at 15 +/- 1º C. The FPSS has been redesigned because the power needed to dissipate is higher than that for the ASTRI-Horn camera. The number of TECs has been increased to 12 units, and they are uniformly deployed under the FPSS. Moreover, the new design allows the PDM assembly from the top to make the maintenance of the focal plane easy.

### 2. Calibration system

The ASTRI-Horn camera has an embedded calibration system composed of a blue laser pulse generator, which injects the light in a side glow optical fiber placed around the perimeter of the entrance window of the camera. When the lids are closed, the diffused and scattered light by the window flashes the PDMs of the focal plane allowing the relative gain calibration of the SiPM (Impiombato et al., 2017). The Calibration System has been updated for the ASTRI Mini-Array camera with the installation of a green laser pulse generator. In the new configuration, both blue and green lasers can flash the PDMs to perform new calibration procedures.

### 3. Conclusions

The ASTRI-Horn dual-mirror aplanatic Cherenkov telescope prototype has been implemented in Sicily on the Mount Etna slope and is now in operation. It will perform scientific programs with its innovative prototype camera based on SiPM sensors. Based on the prototype design, cameras with a more consolidated design are being developed for the ASTRI Mini-Array Gamma ray astronomy experiment being installed in Tenerife, Canary Islands with an unprecedented field of view (10.5°). In this respect, the engineering camera, designed based on the ASTRI-Horn layout, is being developed by CAEN EIE GROUP industrial consortium under the INAF's supervision, aiming at achieving the best performance. The electronic components of the engineering camera have already been produced and are currently under test.

The mechanical components are also in production. The integration and assembly phase will start at the end of January 2023. It is planned to install the engineering camera on the first telescope of the ASTRI Mini-Array at the end of April 2023.

## Acknowledgments
This work was conducted in the context of the ASTRI Project thanks to the support of the Italian Ministry of University and Research (MUR) as well as the Ministry for Economic Development (MISE), with funds explicitly assigned to the Italian National Institute of Astrophysics (INAF). We acknowledge the support of the of Brazilian Funding Agency FAPESP (Grant 2013/10559-5) and the South African Department of Science and Technology through Funding Agreement 0227/2014 for the South African Gamma-Ray Astronomy Program.

IAC is supported by the Spanish Ministry of Science and Innovation (MICIU). They are partially supported by H2020-ASTERICS, a project funded by the European Commission Framework Programme Horizon 2020 Research and Innovation action under grant agreement n. 653477. The ASTRI project is becoming a reality thanks to Giovanni "Nanni" Bignami, Nicolo "Nichi" D'Amico, two outstanding scientists who, in their capability as INAF Presidents, provided continuous support and invaluable guidance. While Nanni was instrumental in starting the ASTRI telescope, Nichi transformed it into the Mini Array in Tenerife. Now the project is being built owing to the unfaltering support of Marco Tavani, the current INAF President. Paolo Vettolani and Filippo Zerbi, the past and current INAF Science Directors, and Massimo Cappi, the Coordinator of the High Energy branch of INAF, have been also very supportive of our work. We are very grateful to all of them. Unfortunately, Nanni and Nichi passed away, but their vision still guides us. This article has gone through the internal ASTRI review process.